\documentclass[aps,prl,twocolumn,showpacs]{revtex4}
\usepackage[dvips]{graphicx}
\usepackage{epstopdf}
\usepackage{indentfirst}
\usepackage{braket}
\usepackage{float}
\usepackage{CJK}
\usepackage{esint}
\usepackage{color}
\usepackage{subfigure}
\usepackage{amsfonts}

\def\be{\begin{equation}}
\def\ee{\end{equation}}
\def\ba{\begin{eqnarray}}
\def\ea{\end{eqnarray}}

\begin{document}
\begin{CJK*}{GBK}{song}
\title{Quantum  thermalization and equilibrium state with multiple temperatures}
\author{Quntao Zhuang\footnote{Now at Department of Physics, Massachusetts Institute of Technology,USA}}
\affiliation{International Center for Quantum Materials, Peking University, 100871, Beijing, China}
\author{Biao Wu}
\affiliation{International Center for Quantum Materials, Peking University, 100871, Beijing, China}
\affiliation{Collaborative Innovation Center of Quantum Matter, Beijing, China}

\begin{abstract}
A large class of isolated quantum system in a pure state can equilibrate and serve as a heat bath.
We show that once the equilibrium is reached, any of its subsystems that is much 
smaller than the isolated system  is thermalized such that the subsystem is governed by the
Gibbs distribution. Within this theoretical framework,  the celebrated superposition principle
of quantum mechanics leads to a prediction of  a thermalized subsystem
with multiple temperatures when the isolated system is in a superposition
state of  energy eigenstates of multiple distinct energy scales.
This multiple-temperature state is at equilibrium, completely different from
a non-equilibrium state that has multiple temperatures at different parts.
Feasible experimental schemes to verify this prediction are discussed.
\end{abstract}
\date{\today}
\maketitle
\end{CJK*}

Standard textbook always starts its discussion of quantum statistical mechanics
with the microcanonical ensemble where every energy eigenstate
in a narrow energy range is equally possible. This ensemble is
usually established with two postulates, equal {\it a priori} probability
and random phases, which are either explicitly or inexplicitly stated in standard 
textbooks~\cite{huang_huang_1987,landau_landau_1958}. 
Interestingly, von Neumann laid down a very different foundation for quantum 
statistical physics in a 1929 paper~\cite{eV}, where he proved both the quantum ergordic theorem 
and the quantum H-theorem  ``in full rigor and without
disorder assumptions." For some reasons, von Neumann's work had been 
forgotten for a long time~\cite{goldstein}; all textbooks start 
with ``disorder assumptions" (or postulates) .

There have been renewed interests in the foundation of quantum statistical mechanics~\cite{lebowitz_boltzmanns_1993,gemmer_quantum_2009,deutsch_quantum_1991,srednicki_chaos_1994,reimann_foundation_2008,rigol_thermalization_2008,biroli_effect_2010,cassidy_generalized_2011,rigol_alternatives_2012,short_equilibration_2011,short_quantum_2012, popescu_entanglement_2006,popescu_foundations_2005,sun2007,Wang2012,
linden_quantum_2009,cho_emergence_2010,ikeda_eigenstate_2011,rigol_relaxation_2007,
reimann_equilibration_2012,xiong_universal_2011,goldstein_canonical_2006,santos_weak_2012,reimann_typicality_2007,yukav_equilibration_2011, yukav_decoherence_2012, yukav_nonequilibrium_2011,riera_thermalization_2012,gogolin_absence_2011,
christian_pure_2010,sho_2012,emerson}. 
Many physicists now agree with von Neumann  that
thermodynamics can be derived from the dynamics of a truly isolated quantum system
without the postulates. These  increased efforts to address the issue of equilibration of an
isolated quantum system are partly due to that an isolated quantum system can now 
be achieved and sustained experimentally for a reasonable long time in a highly 
excited state with ultra-cold atoms~\cite{kinoshita_quantum_2006}.
Great progress has been made.  However, there is much more to be desired.  
For example, will the improved understanding on the foundation of quantum statistics result in new physics?

The quantum ergodic theorem proved by von Neumann is mathematically an inequality~\cite{eV}. 
A different version of this inequality, which is more practical and well defined, 
was proved by Reimann~\cite{reimann_foundation_2008}.  According to these two inequalities 
(or quantum ergodic theorem), these quantum systems
will equilibrate in the sense that fluctuations are very small almost all the time.
Both inequalities can only be applied to systems where there are no degenerate energy-gaps. 
In this work we first show that this quantum ergodic theorem may
be applied to a broader class of systems, which include, for example, quantum chaotic 
systems~\cite{gutzwiller_chaos_1990,berry_regular_1977,berry_semi-classical_1977}. 
This is done by example. We numerically study the dynamics of a quantum chaotic system, 
the Henon-Heiles system~\cite{feit_wave_1984},  which has no bound states and does not satisfy the non-degenerate energy-gap condition established by von Neumann
and Reimann. Nevertheless, we find that
the Henon-Heiles system still equilibrates in  the sense of small fluctuations. Furthermore, 
our numerical results show that  the equilibration is also accompanied by an entropy 
approaching maximum.   This is in agreement in spirit with von Neumann's
quantum H-theorem~\cite{eV}.

We then prove analytically that a subsystem of an isolated quantum system
at equilibrium is thermalized such that it is described by
the Gibbs distribution. Here we distinguish between
equilibration and thermalization:  a system equilibrates if its overall features
no longer change with time while it still evolves  microscopically.
Thermalization is only for a subsystem that  is described by the Gibbs
distribution. A thermalized system
must be at equilibrium but not vice versa.

A natural and surprising outcome of this theoretical framework is that a subsystem
can thermalize with multiple distinct temperatures. This can happen when
the isolated system is in a superposition of energy eigenstates that
concentrate around different energy scales. This thermalized system with
multiple temperatures appears unavoidable for  two reasons. (1) According to 
the quantum ergodic theorem~\cite{eV,reimann_foundation_2008}, the equilibrated
state has the same energy distribution as the initial state. One can manipulate the
energy distribution by choosing a suitable initial condition.   (2) There is no {\it a priori}
reason that an initial state must be in a state which is composed only
of energy eigenstates from a  narrow energy range.  We emphasize that
this multi-temperature state is at equilibrium where both hot and cold exist
in one system: ({\it i}) it is completely different
from a state that is out of equilibrium and has different temperatures at
different parts of the system or for its different degrees of freedom. ({\it ii}) It is also
different from an ensemble of systems where some systems have higher temperatures
while others have lower temperatures. We discuss feasible experimental schemes
with ultra-cold atoms and nuclear spins
to confirm our predication.

\begin{figure*}
\centering
\includegraphics[width=0.95\textwidth]{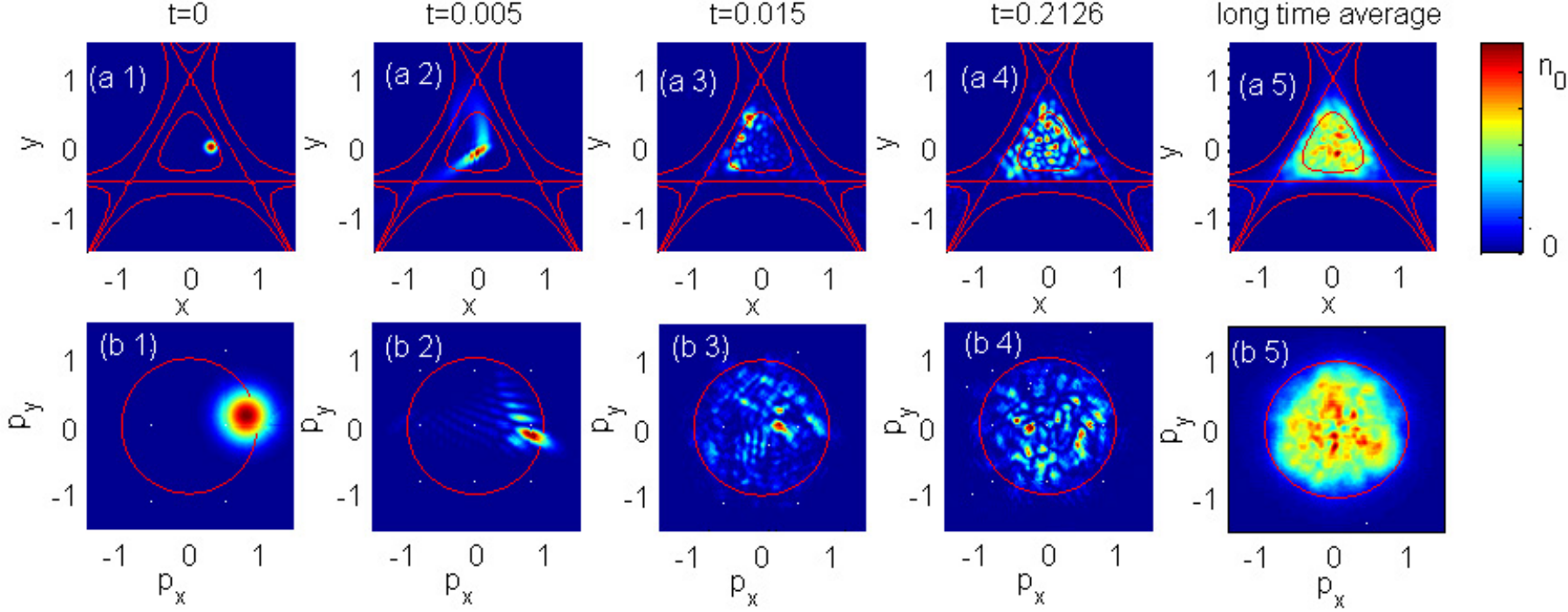}
\caption{{\bf Time evolution of a wave-packet  and the long-time average in the Henon-Heiles
system.} The first row is the density in the real space and the second row is the density
in the momentum space. The long-time averages of these densities are shown at the rightmost
panels. The average is taken over 1200 states equally separated in the time interval of $[0.2012,0.2255]$.
The unit for the real space is $r_c$ and the unit for the momentum space is $p_0$.
The red lines in the first row are energy contours of the Henon-Heiles potential at $V(x,y)/V_c=1/2,~1,~2$.
The red lines in the second row are the maximal classically allowed momentum for the
initial energy of the Gaussian wave packet. The color bars are given on the right side.
\label{hh}
}
\end{figure*}

{\bf Equilibration of an isolated quantum chaotic system.}
The inequality proved for the quantum ergodic theorem by  Reimann~\cite{reimann_foundation_2008} was  slightly modified by
Short {\it et al.}~\cite{short_equilibration_2011,short_quantum_2012}.
This inequality for an observable $A$ now reads
\begin{equation}
\sigma^2_A\equiv \frac{\braket{| \mbox{tr}\{A\rho(t)\}- \mbox{tr}(A\rho_\infty)|^2}_t}{\|A\|^2}\le \frac{1}{d_{\mbox{eff}}}\,,
\label{ineq}
\end{equation}
where $\rho(t)=\ket{\psi(t)}\bra{\psi(t)}$ with $\ket{\psi(t)}=\sum_k c_k\ket{E_k}$ being
the wave function of  the isolated system and $\rho_\infty=\sum_k |c_k|^2\ket{E_k}\bra{E_k}$.  
$\ket{E_k}$ is the energy eigenstate of the system. 
The subscript $t$ in $\braket{}_t$ represents an averaging over a time period much longer than the
characteristic time scale of the system. $\|A\|^2$ is the maximal value of $A$ regarding all the 
states in the Hilbert space. The effective dimension 
$d_{\mbox{eff}}=\frac{1}{\sum_k (\mbox{tr}\ket{E_k}\bra{E_k}\rho(t))^2}$ 
indicates how widely the state $\ket{\Psi}$ is spread over the energy eigenstates.
This inequality holds for a large class of quantum systems that satisfies
the non-degenerate energy-gap conditionin~\cite{eV,reimann_foundation_2008}. 

If a quantum system is in a typical state of high energy, then $d_{\mbox{eff}}$ should
be large since the density of states usually increases tremendously with energy. This means
that the right hand side of Eq.(\ref{ineq}) is small.  Therefore, this inequality tells us
two things: ({\it i}) An isolated quantum system in a high-energy state
will eventually relax to a state where an observable will fluctuate in small amplitude
around its averaged value.  ({\it ii}) Although the isolated quantum system
is described by a wave function,  the expectation
value of any observable $A$ at almost any moment can be computed with
$\rho_\infty$, that is,  $\mbox{tr}(\rho A)\approx\mbox{tr}(\rho_\infty A)$. 

Remarks are warranted here. (1) $\rho_\infty$ is different from the standard micro-canonical
density matrix in textbooks~\cite{huang_huang_1987,landau_landau_1958}:  
the coefficients $|c_k|^2$'s  are determined by the initial condition and they are not necessarily 
equal to each other and distributed in a narrow energy range. (2) The coefficients $|c_k|^2$'s
do not change with time; therefore, the energy distribution of the final equilibrated 
state is the same as the initial state.

For many physicists, small fluctuations already imply equilibrium; however, for others,
a state is equilibrated only when its entropy is maximized.  For the latter group,
even though ground states and other eigenstates  have no fluctuations, they can not 
be regarded as equilibrium.  Von Neumann belongs to the latter group.  
By introducing an entropy for a pure state~\cite{eV},  he proved the quantum H-theorem , 
which demands that a low-entropy state evolve into a high-entropy state.
We address this entropy issue with an example by defining a special 
entropy for pure states in  the single particle Henon-Heiles system~\cite{feit_wave_1984}. 
This is in spirit similar to the entropy
for a pure state introduced by von Neumann, for which there is no known practical 
procedure to compute so far.  Note that this entropy for a pure state introduced 
by von Neumann in 1929 is different from the usual von Neumann 
entropy, which is zero for all pure states. 

We emphasize that it is reasonable to use a single-particle quantum chaotic system
for illustration.  When expressed in the form of matrix, there is no essential difference 
between one-body Hamiltonian and many-body Hamiltonian as long as they 
belong to the same class of random matrix~\cite{gutzwiller_chaos_1990}. 
This is particularly true for the system's energy spectrum, which appears to be
the only factor in determining whether the system equilibrates 
or not~\cite{eV,reimann_foundation_2008}.  We expect that the one-body and 
many-body systems share many dynamical features when their 
corresponding matrices belong to the same class.  More discussion on this point
can be found in Ref.\cite{zhuang2}.

The Hamiltonian of the system for the Henon-Heiles system is $H={p^2}/{2m}+\frac{\alpha}{2}(x^2+y^2)+\lambda (x^2y-\frac{y^3}{3})$,
which has three saddle points located at a distance
$r_c\equiv\frac{\alpha}{\lambda}$ from the origin. These three points  are the
corners of the energy triangular contour with potential
$V_c\equiv\frac{\alpha^3}{6\lambda^2}$ as shown in Fig.\ref{hh}(a1-a5).
The momentum corresponding to the saddle point energy is $p_0\equiv \sqrt{2mV_c}$ as
indicated by the circle in Fig.\ref{hh}(b1-b5).  In our numerical simulation we
set $m=\frac{1}{2}$, $\hbar=1$, and $\alpha/\lambda=1/3$.

The initial condition is a highly localized Gaussian wave packet as shown in Fig.\ref{hh} (a1,b1)
so that the system energy is high. This wave packet is centered at $\vec{r}_i=(0.3,0)r_c$ and $\vec{p}_i=(\cos10^\circ, \sin10^\circ)\sqrt{7/10}p_0$ in the real and momentum spaces, respectively.
A classical particle with $\vec{r}_i$ and $\vec{p}_i$
has energy $0.9691V_c$ and its motion  is fully chaotic.

We numerically solve the  Schr\"{o}dinger equation
and the dynamical evolution of the wave packet is illustrated in
Fig.\ref{hh}(a1-a4,b1-b4). As the wave packet evolves, it begins to
spread out and distort in shape. Eventually it reaches an equilibrium state,
where the wave packet spreads out all over
 the classically allowed region in the real space and the momentum space.
 This overall feature will no longer
change even though the details
of the wave packet still change in the following dynamical evolution.
For comparison, we have  calculated $n_\infty(\vec{r})=\braket{\vec{r}|\rho_\infty|\vec{r}}$ and $n_\infty(\vec{p})=\braket{\vec{p}|\rho_\infty|\vec{p}}$ by long-time averaging, i.e.,
the equilibrium state obtained by Reimann~\cite{reimann_foundation_2008}.
The results are shown  in Fig.\ref{hh}(a5,b5).
It is clear that $n_\infty(\vec{r})$ and $n_\infty(\vec{p})$ are very similar to
the wave packet at $t=0.2126$ with the same overall feature that the wave function
spread all over both the triangular spatial region and the circled momentum region except for some fluctuations.

\begin{figure}
\centering
\includegraphics[width=0.45\textwidth]{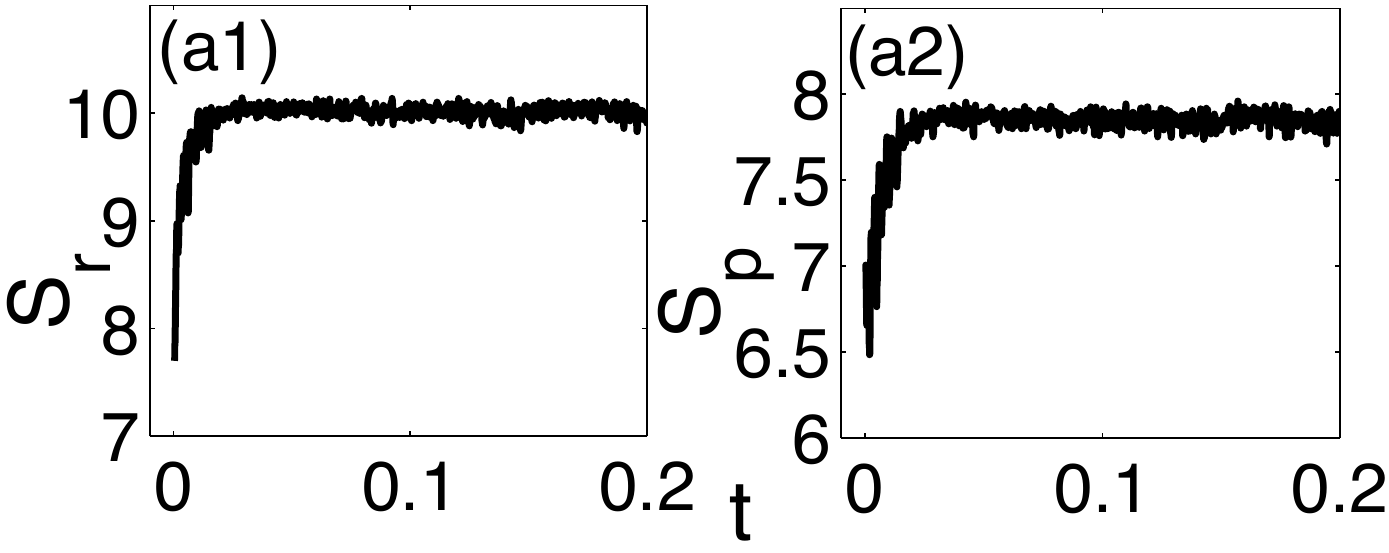}
\caption{ {\bf 
Time evolution of the entropies
$S_\xi$ for the Henon-Heiles system for both spatial and momentum space.}
\label{entropy}
}
\end{figure}

To demonstrate the system has  relaxed to an equilibrium state,  we 
define an entropy as ${S_\xi}=-\int d\xi\frac{n(\xi,t)}{n_\infty(\xi)}\ln(\frac{n(\xi,t)}{n_\infty(\xi)}), \xi=\vec{r},~\vec{p}$.  This entropy indicates how wide spread the wave function is in the classically-allowed
region. The time evolution of ${S_\xi}$ is shown  in Fig.\ref{entropy}(a1,a2),  where we see
clearly the entropies quickly saturate and reach the maximum values, indicating that an
equilibrium state is reached. Note  that the relaxation times
in both the real and momentum spaces are about the same. However, it must be
pointed out that this definition of entropy applies only for some special systems and
do not apply for a general quantum system. It is
in spirit in accordance with the entropy for a pure state introduced for  a general system 
 by von Neumann~\cite{eV}. 

The equilibrium state reached is consistent with the inequality Eq.(\ref{ineq}).
 To check the inequality numerically, one needs to compute
 energy  eigenstates of the system.  As it is difficult to compute the eigenstates for the
Henon-Heiles system, we have turned to the ripple billiard system studied in Ref.\cite{xiong_universal_2011,li_quantum_2002} to verify the inequality. 
The verification is successful and the detailed computation 
can found in Ref.\cite{zhuang2}. We only mention here that $d_{\rm eff}\approx 300$
for a similar Gaussian wave packet in the ripple billiard system. 

Note that the quantum ergodic theorem was originally proved by von Neumann 
and Reimann for systems that have no degenerate energy gaps~\cite{eV,reimann_foundation_2008}. 
It was later generalized to systems that have limited amount of degeneracy
~\cite{short_quantum_2012}. However, it is still not clear that how these degeneracy conditions
are related to the integrability of the systems. Our numerical simulations here 
and elsewhere \cite{xiong_universal_2011,zhuang2} suggest that this theorem  
may be applied in a much broader class of quantum systems, which include quantum chaotic systems.

{\bf Thermalization of subsystems.}
We have shown that a large class of truly isolated quantum systems, including chaotic systems,  can relax to
an equilibrium state. Now we decompose an equilibrated isolated quantum 
system into two parts, subsystem $S$ and thermal bath $B$.
We consider an operator $\mbox{tr}^{(B)}$, which traces out the thermal
bath $B$ and gives the density operator for the small subsystem $S$.
Note the subsystem $S$ is small compared with
system $S+B$ but still large on the microscopic level. Based on our equilibration
picture, the expectation value of $\mbox{tr}^{(B)}$ should also equilibrate.
We shall show that due to the coupling to the rest of the system,
these equilibrated subsystems are also thermalized
so that they are described by the Gibbs distribution. The derivation of Gibbs distribution
for a subsystem has been considered before with the assumption that the isolated system is in
a pure state composed of energy eigenstates from a small energy interval~\cite{reimann_typicality_2007,riera_thermalization_2012}.  We show that this assumption
is not necessary and when the pure state is composed of energy
eigenstates of different energy scales, the subsystem is thermalized  with
multiple temperatures.

We write the Hamiltonian of the isolated system as $H^{S+B}=H^S+H^B+\Delta H$,
where $\Delta H$ is the weak interaction between system $H^S$ and thermal bath $H^B$.
Suppose that the composite system is described by a wave function $\ket{\Phi^{S+B}}=\sum c_k \ket{E^{S+B}_k}$,  where $\ket{E^{S+B}_k}$'s are the energy eigenstates
of the composite system. By tracing out thermal bath $B$, we obtain the density
operator for system $S$,
$\rho^S=\mbox{tr}^{(B)}\ket{\Phi^{S+B}}\bra{\Phi^{S+B}}$.
The system will eventually equilibrate; as an observable, $\rho^S$ will
be close to its long time average, i.e.
$\rho^S\approx\braket{\rho^S}_t\equiv\mbox{tr}^{(B)}\rho_\infty^{S+B}=\mbox{tr}^{(B)}\sum |c_k|^2\ket{E^{S+B}_k}\bra{E^{S+B}_k}$.

We  expand the eigenstate $\ket{E^{S+B}_k}$ as follows
\begin{equation}
\ket{E_k^{S+B}}\approx {\sum}^\prime a_{ij}^k \ket{E^S_i}\ket{E^B_j}\,,
\label{approx}
\end{equation}
where $\ket{E^S_i}$ and $\ket{E^B_j}$ are energy eigenstates of system $S$ and thermal bath $B$,
respectively. The prime above indicates that the summation is only over eigen-energies satisfying
\begin{equation}
E^{S+B}_k=E^S_i+E^B_j+\Delta E_{ij}\,.
\label{comb}
\end{equation}
where $\Delta E_{ij}$ is the interaction energy that is usually very small compared to
$E^S_i$ and $E^B_j$ when long-range interaction is negligible, e.g., gravity, in the system.
Two remarks are warranted. ({\it i}) The approximation made in Eq.(\ref{approx}) is justified.
The equality holds when there is no coupling $\Delta H=0$. We expect it hold when the weak
interaction $\Delta H$ is turned on.  ({\it ii}) The weak interaction $\Delta H$ can drive the
system to a state with $a^k_{ij}$'s randomly uniformly distributed on the sphere
$\sum_{ij}|a^k_{ij}|^2=1$.
This random distribution is similar to the idea of  "typicality"~\cite{goldstein_canonical_2006,santos_weak_2012}. The connection between
interaction and randomness is widely acknowledged since the details of
the interaction is irrelevant to the statistical properties~\cite{flambaum_towards_1996,borgonovi_chaos_1998}.
As a result, the average value of $|a^k_{ij}|^2$ is $\frac{1}{D^{S+B}(E^{S+B}_k)}$,  where
$D^{S+B}(E^{S+B}_k)$ is the degeneracy brought by the combination of states.  We emphasize
that this degeneracy is different from the intrinsic degeneracy of the system and it is due to
the existence of $\Delta E_{ij}$ in Eq.(\ref{comb}) .

With the approximation made in Eq.(\ref{approx}), we now proceed with our derivation,
\ba
\rho^S&=&\sum_k |c_k|^2 \sum_m \braket{E_m^B| E^{S+B}_k}\braket{E^{S+B}_{k}|E_m^B}
\nonumber\\
&=&\sum_k |c_k|^2 \sum_{i i^\prime} {}^\prime \{\sum_m a^k_{im}a^{k*}_{i^\prime m}\}\ket{E^S_i}\bra{E^S_{i^\prime}}\nonumber\\
&=&\sum_k |c_k|^2 \sum_i (\sum_m |a^k_{im}|^2)\ket{E^S_i}\bra{E^S_{i}}\nonumber\\
&&+\sum_k |c_k|^2 \sum_{i\neq i^\prime}^\prime(\sum_m a^k_{im}a^{k*}_{i^\prime m})\ket{E^S_i}\bra{E^S_{i^\prime}}\,.
\label{derivation}
\ea
The central limit theorem gives the results of the first summation as $\sum_m |a^k_{im}|^2\simeq D^B(E^B_m)/D^{S+B}(E^{S+B}_k)$ and the second summation as
$\sum_m a^k_{im}a^{k*}_{i^\prime m}\sim O\{\sqrt{D^B(E^B_m)}/D^{S+B}(E^{S+B}_k)\}\simeq O\{1/\sqrt{D^{S+B}(E^{S+B}_k)}\}$
where $D^B(E^B_m)$ is the degeneracy of the thermal bath and we have used that $S$ is much smaller than $B$ so that
$D^B(E^B_m)\simeq D^{S+B}(E^{S+B}_k)$.

As a result, the last term in Eq.(\ref{derivation})
has the  order of magnitude at $O\{ \sum_i{D^{S}(E_i)}^2/\sqrt{D^{S+B}(E^{S+B}_k)}\}$,
which is practically zero for the isolated system is much larger than system $S$.
So omitting the last term, we have from Eq.(\ref{derivation})
\begin{equation}
\rho^S=\sum_k |c_k|^2 \sum_i \frac{D^B(E^B_m)}{D^{S+B}(E^{S+B}_k)}
\ket{E^S_i}\bra{E^S_{i}}\,.
\end{equation}
With the standard argument for the Gibbs distribution~\cite{landau_landau_1958},
we arrive finally at
\be
\rho^S= \sum_k |c_k|^2\{\sum_i \exp(-\beta_k E_i)\ket{E^S_i}\bra{E^S_{i}}\}\,,
\label{final}
\ee
where $\beta_k\equiv\frac{1}{k_BT_k}\equiv{\partial\ln D^{S+B}(E^{S+B}_k)}/{\partial E^{S+B}_k}$ defines
the temperature of the total system for eigenstate $\ket{\Phi_k^{S+B}}$. In this way,
we have proved that a subsystem of an isolated equilibrated system is thermalized.

{\bf Thermalized state with multiple temperatures.}
Here we examine Eq. (\ref{final}) for two typical cases: ({\it i}) The coefficients $|c_k|^2$ of the composite system have a single sharp peak distribution around energy $E_{p}$.
This case is considered by others
~\cite{reimann_typicality_2007} in different formalisms.  For this case,  the density matrix
$\rho^S$ in Eq.(\ref{final}) is reduced to $\rho^S=\sum_i \exp(-\beta_p E_i)\ket{\psi^S_i}\bra{\psi^S_{i}}$.
This is exactly the typical Gibbs distribution discussed in all textbook on statistical mechanics.
({\it ii}) The coefficients $|c_k|^2$ have two well-separated sharp peaks  around two energies
$E_{p1}$ and $E_{p2}$. In other words, the composite system (or the heat bath) is in a superposition of {\it numerous} eigenstates centered around two very different energy scales.
In this case, the thermalized system has two temperatures, $\beta_1$ for $E_{p1}$
and $\beta_2$ for $E_{p2}$, with the following density matrix
\be
\rho^S= \sum_i (|a_1|^2e^{-\beta_1 E_i}+|a_2|^2e^{-\beta_2 E_i})\ket{E^S_i}\bra{E^S_{i}}\,,
\label{bite}
\ee
where $|a_1|^2$ and $|a_2|^2$ are the weight of the two distribution peaks.
The following is a list of key points for a good understanding of this quantum equilibrium state
with two different temperatures. 

\renewcommand{\labelenumi}{(\alph{enumi})}
\begin{enumerate}
\item When the quantum heat bath 
is in a superposition of states with two well-separated energy scales, each particle 
in the subsystem always feel different energy scales simultaneously when it exchanges
energy with the  heat bath. This leads to a thermalized state with two different temperatures.

\item When a system is in such a state, it consists of two parts, one hot and one cold.
However, one can not tell which particle belongs to the hot part and which particle
is in the cold part. This is similar to liquid helium.  It consists of a superfluid part
and a normal fluid part; but no single helium atom can be assigned to either the
superfluid part or the normal fluid part. 

\item When an ideal gas is thermalized to such an equilibrium state with two temperatures, 
each particle in the gas can be roughly viewed  as in  a superposition state of 
two different momenta. This is impossible in a classical ideal gas, where each particle has
a definite momentum. 

\item Since the total
system $S+B$ is isolated, the coefficients $|c_k|^2$s are constants of motion and only
depend on the initial condition. As a result,  the two peaks in the initial distribution of $|c_k|^2$'s
will remain intact during the whole dynamical process.  In other words, the system is stable
with the double-peak energy distribution. 

\item If the total system $S+B$ is a superposition of just two different energy eigenstates, 
the total system is not in an equilibrium state. This is because in this case $d_{\rm eft}=2$ and
the left hand side of the inequality is large. To ensure equilibration,  we need $d_{\rm eft}\gg 1$,
that is to  have large number of eigenstates concentrating around two different energy scales. 

\item This state does not describe a statistical ensemble
of systems, where some systems are cold and some systems are hot.

\item  Our state is an equilibrium state with multiple
temperatures; it  is completely different from the usual non-equilibrium
state which  has different temperatures for different parts. 

\item  Our state is not a Schr\"odinger cat state~\cite{Cat};  it does not collapse upon  measurement. 
\end{enumerate}

For most of  the systems that we have encountered in nature or studied in experiment, they
are in contact with classical heat bath. However, with the advance of technology,
we can now create large quantum systems which can serve as quantum heat bath.
Two such examples are Bose-Einstein condensates (BECs) and nuclear spins in a
quantum dot, where feasible experiments can be set up to test our prediction.
({\it i}) Consider a two-species BEC. One species with larger population
is trapped  optically  in an uneven double-well
potential~\cite{shin_distillation_2004} while  the smaller species is trapped in a single well potential.
The larger species serves as a heat bath with  two energy peaks due to uneven double-well
potential. By exchanging energy with the larger species,
the smaller species should develop a double-peak
distribution in momentum space, signaling the existence of two temperatures.
Since the uneven double-well has to be kept for the double-peaked energy distribution,
the state realized here is not strictly equilibrium and might be more accurately called stationary.
Now a two-species BEC has been realized just recently in experiment~\cite{wangdj}
({\it ii}) Due to the weak coupling to the enviornment,  nuclear spins in a quantum dot
can be regarded as quantum bath for a long time~\cite{yao_prb_2006,zhao_anomalous_2011,huang_observation_2011}.
With the feedback technique that has been demonstrated both theoretically and experimentally
\cite{yao_nature},
one should be readily design a scheme that can put these nuclear spins in a superposition
state of two different energy scales and use the electron spin to probe such a state~\cite{yao_private}.

Note that Fine {\it et al.} have also abandoned the transitional
microcanonical ensemble and replaced it with ``quantum microcanonical" ensemble~\cite{Fine3,Fine2,Fine1}.
This is fundamentally different from our approach,
where no assumption for an ensemble is needed.


\begin{acknowledgments}
We acknowledge helpful discussion with Hongwei Xiong , Qian Niu, and Wang Yao.
This work is supported by the NBRP of China (2012CB921300,2013CB921900) and
the NSF of China (10825417,11274024,11334001) and the RFDP of China (20110001110091).
\end{acknowledgments}


\begin{thebibliography}{35}
\bibitem{huang_huang_1987}
K. Huang,
 {Statistical Mechanics}, page 176. John Wiley \& Sons,
  Inc., 2 edition (1987).

\bibitem{landau_landau_1958}
L. Landau and E. Lifshitz,
 {Statistical Physics}, pages 78--80. Pergamon, London(1958).
 
\bibitem{eV}
J. von Neumann, Z. Phys. 57, 30 (1929); [Eur. Phys. J. H
{\bf 35}, 201 (2010)].

\bibitem{goldstein}
S. Goldstein,  J. L. Lebowitz, R. Tumulka,  and N. Zanghi
European Phys. J. H {\bf 35}, 173 (2010).

\bibitem{lebowitz_boltzmanns_1993}
J.L. Lebowitz,
 {Phys. Today}, 46(9), 32 (1993).

\bibitem{gemmer_quantum_2009}
J. Gemmer, M. Michel, and G. Mahler,
 {Quantum Thermodynamics}, Lecture Notes in Physics, pages
  86--126. Springer--Verlag Berlin Heidelberg, 2 edition (2009).
  
\bibitem{deutsch_quantum_1991}
J.M. Deutsch,
 {Phys. Rev. A}, 43, 2046 (1991).
 
\bibitem{srednicki_chaos_1994}
M. Srednicki, {Phys. Rev. E}, 50, 888 (1994).

\bibitem{reimann_foundation_2008}
P. Reimann, {Phys. Rev. Lett.}, 101, 190403 (2008).

\bibitem{rigol_thermalization_2008}
M. Rigol, V. Dunjko, and M. Olshanii,
 {Nature}, 452, 854 (2008).

\bibitem{biroli_effect_2010}
G. Biroli, C. Kollath, and A.M. L\"{a}uchli,
 {Phys. Rev. Lett.}, 105, 250401 (2010).

\bibitem{cassidy_generalized_2011}
A.C. Cassidy, C.W. Clark, and M. Rigol,
 {Phys. Rev. Lett.}, 106, 140405 (2011).


\bibitem{rigol_alternatives_2012}
M. Rigol and M. Srednicki,
 {Phys. Rev. Lett.}, 108, 110601 (2012).


\bibitem{short_equilibration_2011}
A.J. Short,
 {New J. Phys.}, 13(5), 053009 (2011).

\bibitem{short_quantum_2012}
A.J. Short and T.C. Farrelly,
 {New J. Phys.}, 14, 013063 (2012).

\bibitem{popescu_entanglement_2006}
S. Popescu, A.J. Short, and A. Winter,
 {Nature Phys.}, 2, 754 (2006).

\bibitem{popescu_foundations_2005}
S. Popescu, A.J. Short, and A. Winter,
 {arXiv:quant-ph/0511225} (2005).
 
\bibitem{sun2007}H. Dong, S. Yang, X. F. Liu, and C. P. Sun, 
Phys. Rev. A {\bf 76}, 044104 (2007).

\bibitem{Wang2012}W. Wang, Phys. Rev. E {\bf 86}, 011115 (2012).

\bibitem{linden_quantum_2009}
N. Linden, S. Popescu, A.J. Short, and A. Winter,
 {Phys. Rev. E}, 79, 061103 (2009).

\bibitem{cho_emergence_2010}
J. Cho and M.S. Kim,
 {Phys. Rev. Lett.}, 104, 170402 (2010).

\bibitem{ikeda_eigenstate_2011}
T.N. Ikeda, Y. Watanabe, and M. Ueda,
 {Phys. Rev. E}, 84, 021130 (2011).

\bibitem{rigol_relaxation_2007}
M. Rigol, V. Dunjko, V. Yurovsky, and M. Olshanii,
  {Phys. Rev. Lett.}, 98, 050405 (2007).



\bibitem{reimann_equilibration_2012}
P. Reimann and M. Kastner,
 {New J. Phys.}, 14, 043020 (2012).

\bibitem{xiong_universal_2011}
H. Xiong and B. Wu,
 {Laser Phys. Lett.}, 8, 398 (2011).

\bibitem{goldstein_canonical_2006}
S. Goldstein, J.L. Lebowitz, R. Tumulka, and N. Zangh\'{i},
 {Phys. Rev. Lett.}, 96, 050403 (2006).


\bibitem{santos_weak_2012}
L.F. Santos, A. Polkovnikov, and M. Rigol,
 {Phys. Rev. E}, 86, 010102 (2012).


\bibitem{reimann_typicality_2007}
P. Reimann,
 {Phys. Rev. Lett.}, 99, 160404 (2007).

\bibitem{yukav_equilibration_2011}
V.I. Yukalov,
 {Laser Phys. Lett.}, 8, 485 (2011).


\bibitem{yukav_decoherence_2012}
V.I. Yukalov,
 {Ann. Phys.}, 327, 253 (2012).


\bibitem{yukav_nonequilibrium_2011}
V.I. Yukalov,
 {Phys. Lett. A}, 375, 2797 (2011).

\bibitem{riera_thermalization_2012}
A. Riera, C. Gogolin, and J. Eisert,
 {Phys. Rev. Lett.}, 108, 080402 (2012).

\bibitem{gogolin_absence_2011}
C. Gogolin, M.P. M\"uller, and J. Eisert,
 {Phys. Rev. Lett.}, 106(4), 040401 (2011).


\bibitem{christian_pure_2010}
C. Gogolin,
 {{arXiv:quant-ph/1003.5058}} (2010).
\bibitem{sho_2012}
S. Sugiura and A. Shimizu,
 {Phys. Rev. Lett.}, 108, 240401 (2012); S. Sugiura and A. Shimizu 
Phys. Rev. Lett. 111, 010401 (2013).
 
\bibitem{emerson} C. Ududec, N. Wiebe, and J. Emerson, Phys. Rev. Lett.
111, 080403 (2013); S. Goldstein, T. Hara, and H. Tasaki 
Phys. Rev. Lett. 111, 140401 (2013).


\bibitem{kinoshita_quantum_2006}
T. Kinoshita, T. Wenger, and D.S. Weiss,
 {Nature}, 440, 900 (2006).

\bibitem{gutzwiller_chaos_1990}
H-J St\"{o}ckmann,
{ Quantum Chaos, an introduction}, Chapter 2, Cambridge University Press (1999).



\bibitem{berry_regular_1977}
M.V. Berry,
  {J. Phys. A: Math. Gen.}, 10, 2083 (1977).

\bibitem{berry_semi-classical_1977}
M.V. Berry,
  {Phil. Trans. R. Soc. A}, 287, 237 (1977).


\bibitem{feit_wave_1984}
M.D. Feit and J.A. Fleck,
 {J. Chem. Phys.}, 80, 2578 (1984); R. A. Pullen and A. R. Edmonds, J. Phys. A: Math.
Gen. 14 L477 (1981); D Engel, J Main, and G Wunner, J. Phys. A: Math. Gen. 31 (1998) 6965; M. Brack, R. K. Bhaduri, J. Law, and M. V. N. Murthy, Phys. Rev. Lett. 70, 568 (1993); D. W. Noid and R. A. Marcus, J. Chem. Phys. 67, 559 (1977).

\bibitem{zhuang2}
Q. Zhuang and B. Wu, arXiv:1308.1717 (2013). 


\bibitem{li_quantum_2002}W. Li, L.E. Reichl, and B. Wu,
 {Phys. Rev. E}, 65, 056220 (2002).
 



\bibitem{flambaum_towards_1996}
V.V. Flambaum, F.M. Izrailev, and G. Casati,
 {Phys. Rev. E}, 54, 2136 (1996).

\bibitem{borgonovi_chaos_1998}
F. Borgonovi, I. Guarneri, F. Izrailev, and G. Casati,
 {Phys. Lett. A}, 247, 140 (1998).

\bibitem{shin_distillation_2004}
Y. Shin \it{et al.}\rm,
 {Phys. Rev. Lett.}, 92, 150401 (2004).

\bibitem{Cat}
A.J. Leggett,
 {Science}, 307, 871 (2005).
 
\bibitem{wangdj}Dezhi Xiong, Xiaoke Li, Fudong Wang, and Dajun Wang, arXiv:1305.7091 (2013).

\bibitem{yao_prb_2006} W. Yao, R.B. Liu, L.J. Sham,
 {Phys. Rev. B}, 74, 195301 (2006).

\bibitem{zhao_anomalous_2011}
N. Zhao, Z.Y. Wang, and R.B. Liu,
 {Phys. Rev. Lett.}, 106, 217205 (2011).

\bibitem{huang_observation_2011}
P. Huang \it{et al.}\rm,
 {Nature Commu.}, 2, 570 (2011).

\bibitem{yao_nature}
X.-D. Xu \it{et al.}\rm,
 {Nature}, 459, 1105 (2009);
 W. Yao and Y. Luo, EPL, 92 (2010) 17008.

\bibitem{yao_private} Wang Yao, private communication.

\bibitem{Fine1}
K. Ji and B.V. Fine,
 {Phys. Rev. Lett.}, 107, 050401 (2011).

\bibitem{Fine2}
B.V. Fine and F. Hantschel,
 {arXiv:cond-mat/1010.4673} (2010).

\bibitem{Fine3}
B.V. Fine,
 {Phys. Rev. E}, 80, 051130 (2009).

\end{thebibliography}
\end{document}